 \documentclass{article}
 \usepackage{arxiv}
 
\usepackage{amssymb}
\usepackage{amsfonts}
\usepackage{amsmath}
\usepackage{graphicx}%
\usepackage{cite}
\usepackage{xcolor}
\definecolor{brown}{rgb}{0.59,0.29,0}

\usepackage[normalem]{ulem}
\providecommand{\U}[1]{\protect\rule{.1in}{.1in}}
 
\title{ADMM and Spectral Proximity Operators in Hyperspectral Broadband Phase Retrieval for Quantitative Phase Imaging}
\author{Vladimir Katkovnik, Igor Shevkunov, Karen Egiazarian%
\\Computational imaging group, Tampere University, Tampere, Finland, e-mail: igor.shevkunov@tuni.fi }
\begin{document}
\maketitle
\begin{abstract}
A novel formulation of the  hyperspectral
broadband phase retrieval  is developed for the scenario  where both object and modulation phase masks are spectrally varying.  The proposed algorithm is based on a complex domain version of
the  alternating direction method of multipliers (ADMM) and Spectral Proximity
Operators (SPO) derived for Gaussian and Poissonian observations.   Computations for these operators are reduced to  the solution of sets of cubic (for Gaussian)
and quadratic (for Poissonian) algebraic equations. These proximity operators resolve two problems. Firstly, the complex domain
spectral components of signals are extracted from the total intensity observations calculated as  sums of the signal spectral intensities. In this way, 
the spectral analysis of the total intensities is achieved. Secondly, the noisy
observations are filtered, compromising noisy intensity observations and their
predicted counterparts.
The ability to resolve the hyperspectral broadband phase retrieval problem and to find the spectrum varying object are  essentially defined by the spectral properties of object and image formation operators. The simulation tests demonstrate that the phase retrieval in this formulation can be successfully resolved. 
\end{abstract}


\section{Introduction}

{Multispectral (MS) and hyperspectral (HS) images differ in amount and width of spectral
channels. The number of spectral channels varies from a few to
tens  for MS images and goes from tens  up to hundreds  for HS
images. We do not make a difference between these two scenarios and use the word 
'hyperspectral' addressing to both MS and HS ones. The HS imaging is much more informative as compared with
the conventional RGB imaging and indispensable in many applications such as remote sensing, biology,
medicine, quality material and food characterization, control of ocean and
earth pollution, etc. \ A flow of publications on computational methods developed for
various applications of HS imaging is growing fast (e.g. \cite{HS_Review_1,HS_Review_2,HS_Review_3}).

A spectral information is provided by  optical devices (spectrometers) or in numerical form  by  computational methods in two different modes: (1)  registered and  processed channel-by-channel or (2) registered as the total power of entire spectral channels with a subsequent spectral analysis. 
In this paper, we consider the latter scenario with a broadband  illumination of an object of interest and a broadband sensor registering the total power of the impinging light beam.

The HS phase imaging is a comparatively new development dealing
with a phase delay of a coherent light in transparent or reflective objects \cite{Kalenkov2014,Kalenkov2017}.
The
HS broadband phase imaging is  more informative than the monochromatic one and provides more precise
information and  visualization of invisible. It is one of the
most promising  directions in   quantitative phase imaging \cite{Popesku2012, QUANTITATIVE_PHASE_IM_1,Waller2019,QUANTITATIVE_PHASE_IM_3} with numerous applications in optical metrology \cite{Claus2018}, microscopy \cite{Kemper2006}, digital holography\cite{Tahara2018a},
biology and medicine\cite{Baek2020}.  
In particular, the quantitative phase imaging enables  label-free and
quantitative assessment of cells and tissues, which plays an important role in
 study of their optical, chemical, and mechanical intrinsic properties
\cite{BioMedicine,Kemper2018_HS,Ba2018b,Yushkov2020}.

Contrary to the intensity imaging, the phase imaging is
more complex as the quantitative phase cannot be measured directly and should  be extracted from the indirect intensity observations.
The conventional phase imaging employing the principles of interferometry  exploits the reference beams, and the interference between the object and the reference beams is a source of the phase information for the object.

Conventionally, for  the processing of HS images, 2D spectral narrow-band images are
stacked together and represented as 3D cubes with two spatial coordinates
$(x,y)$ and the third longitudinal spectral coordinate.
In phase imaging, data in these 3D cubes are complex-valued with spatial  and
spectral varying amplitudes and phases. Due to this, phase image processing is more complex than the HS intensity imaging where the corresponding 3D cubes are real-valued.

The usual setup of monochromatic phase
retrieval assumes  recovering  of a complex-valued object $U_{o}\in\mathbb{C}%
^{N}$ from  multiple (or single) intensity measurements (squared linear
projections) $Y_{t}=|A_{t}U_{o}|^{2},$ $t=1,...,T$, where $Y_{t}\in
\mathbb{R}^{M}$, and $A_{t}\in\mathbb{C}^{M\times N}$ is an image formation
operator. The heuristic  iterative algorithms with
alternative projections to the object and measurement planes are well known and studied starting 
from the works of Gerchberg and Saxton \cite{Gerchberg1969} and Fienup \cite{Fienup-1982}.
These techniques are proven  to be efficient for various optical applications.

The  common setup  for phase retrieval assumes a modulation of the object
$U_{o}$ by the random phase mask $\mathcal{M}_{t}\in\mathbb{C}^{N}$ with the
observation model
\begin{equation}
Y_{t}=|A_{t}(\mathcal{M}_{t}\circ U_{o})|^{2},\text{ }t=1,...,T,\label{StandardPR}%
\end{equation}
where '$\circ$' stands for the Hadamard,  element-by-element, product of two vectors.

In general, the phase retrieval  is a non-convex inverse imaging problem.  Theoretical studies
formulating restrictions on object and image formation models leading to
uniqueness and convergence of the iterative algorithms are of special
interest. The model \eqref{StandardPR} with the operator $A_{t}$ given as the
Fourier transform (FT) is common for many theoretical works \cite{Candes2008},
\cite{EldarReview}, \cite{Giannakic}. Optimization of the phase coding  modulation masks
$\mathcal{M}_{t}$ is
a hot topics  in phase retrieval \cite{Waller2019},
\cite{HenryMaskDesign},
\cite{Overleaf}.
The extended review of the recent developments in mathematical foundation of the phase retrieval problems can be seen in
\cite{UniqunessStability}, \cite{Fannjang2}, \cite{NONCONVEX}.

In this paper, we propose a novel phase retrieval formulation
providing a prospective powerful instrumentation for broadband HS complex domain
(phase/amplitude) imaging.
In what follows we use the  convenient  vectorized  representation of $2D$ images as  vectors. For the object of interest it is the vector 
$U_{o,k}\in\mathbb{C}^{N}$, $N=nm$, where $n$ and $m$  are width and height of $2D$ image; $k$ stays for the
spectral variable, which is a wavenumber  in optics, $k=\dfrac{2\pi}{\lambda}$,
$\lambda$ is a wavelength.

We introduce the HS broadband phase retrieval as a reconstruction
of the complex-valued object $U_{o,k}\in\mathbb{C}^{N},$ $k\in K$, from
the intensity measurements:
\begin{equation}
Y_{t}=\sum_{k\in K}|A_{t,k}U_{o,k}|^{2},\text{ }t\in T.\label{ourPR}%
\end{equation}
For the noisy intensity observations with additive noise $\varepsilon_{t}$, $Y_{t}$ is replaced by $Z_{t}$:
\begin{equation}
Z_{t}=Y_{t}+\varepsilon_{t}, ~t\in T.
\label{ourPR_noisy}%
\end{equation}
Here  $A_{t,k}\in\mathbb{C}^{M\times N}$ are linear operators modeling
\ a propagation of 2D object images from the object plane to the
sensor, \ and $t$ is the index of experiment.
The total intensity measurements $Y_{t}\in
R_{+}^{M}$  are calculated over the spectrum $K$ as the sum of the spectral intensities
$|A_{t,k}U_{o,k}|^{2}$. It is essential in
this paper, that the object $U_{o,k}$ and the operators $A_{t,k}$ are spectrum
dependent, varying in $k$. In optics, it means that the reflective and
transmissive  properties of the object (specimen) 
as well as the light propagation operators depend on the wavelength of light.

There is a small number of publications relevant to the HS broadband phase retrieval from the total intensity measurements considered in this paper. We briefly review these works.

 The HS phase retrieval in the paper \cite{ProximalOperator-2019} is formulated as object reconstruction from the intensity  observations  $Y_{t}=\sum_{k\in K}|A_{t,k}U_{o}|^{2}$, where the object $U_{o}$ is spectrally invariant, i.e. does not depends on $k$.
This assumption differs  \cite{ProximalOperator-2019} essentially from the setup considered in this paper.

With the introduced observation model \eqref{ourPR}, the interferometric HS
methods also can be interpreted as the phase retrieval problems. Indeed, the HS
digital holography uses the observation model
\begin{equation}
Y_{t}=\sum_{k\in K}|A_{k}(R_{t,k}+U_{o,k})|^{2},\text{ }R_{t.k}=\exp(-j2\pi
tk),\text{ }t\in T,\label{DigHolography}%
\end{equation}
where $R_{t.k}$ stays for the harmonic reference signal varying from experiment-to-experiment
\cite{DigitalHolography}, \cite{KalenkovDH},\cite{shevkunov_CCF_2020}.

For the shearing HS digital holography, we can introduce the observations as
\begin{equation}
Y_{t}=\sum_{k}|A_{k}(1+R_{t,k})U_{o,k}|^{2},\text{ }R_{t.k}=\exp(-j2\pi
tk),t\in T,\label{ShearingHolography}%
\end{equation}
i.e., the reference beam $R_{t,k}$ is not additive to the object as in
\eqref{DigHolography} but propagates through the object $U_{o,k}$
 \cite{Kalenkov_selfref_2019},\cite{Shevkunov_HSPR_2020, CCF-ICIP2020}.


The harmonic modulation of the signals in \eqref{DigHolography} and
\eqref{ShearingHolography} is a special feature in computational
holography. The solution for \eqref{DigHolography} can be given by FT of the observations
combined with complex domain filtering \cite{shevkunov_CCF_2020}. 
The solution
for \eqref{ShearingHolography} is obtained by FT of the observations included in
the phase retrieval iterations \cite{CCF-ICIP2020}.  

Note, that the 
measurements $Y_{t}$ in \eqref{ourPR} with arbitrary $A_{t,k}$ are very different from those in \eqref{DigHolography} and
\eqref{ShearingHolography}, where the harmonic modulation is used for spectral analysis of observations typical for interferometry and holography.   The observations \eqref{ourPR}-\eqref{ourPR_noisy} do not include any spectral analysis tools. The ability to resolve the HS phase retrieval problem in our setup is completely defined by the spectral properties of the object $U_{o,k}$ and the image formation operators   $A_{t,k}$.

 The main novelty of this paper is the formulation of the HS phase retrieval for a spectrally varying object  from intensity observations \eqref{ourPR}-\eqref{ourPR_noisy} with spectrally varying image formation operators. 
 The developed algorithm is derived from the variational formalization of the problem  based on
the alternating direction method of multipliers (ADMM) for complex-valued variables and Spectral Proximity
Operators (SPO) obtained for Gaussian and Poissonian observations.
Computations for these operators are reduced to the solution of the sets of cubic (for Gaussian)
and quadratic (for Poisson) algebraic equations. 
The model of the object $U_{o,k}$ is unknown and does not exploited in the algorithm iterations. 

Pragmatically, we are targeted on the HS phase retrieval formulation and 
the algorithm's development. 

The numerical tests prove that the HS phase retrieval in this
general formulation can be resolved. The proposed setup allows a simple
optical implementation, possibly lensless, which is much simpler than
implementations used in interferometry and holography imaging.

The rest of the paper is organized as follows. The algorithm development is a topic of  Section~\ref{Algorithm development}. From a simple variational formulation of the problem, one fidelity term and one prior term, we go to ADMM  considering the features concerning the formulation for complex-valued variables. The spectral proximity operators are introduced as solutions of optimization problems with quadratic penalization.   A Complex-Domain Block-Matching  filter is exploited as  a regularization tool for the HS phase retrieval.
The results of simulation tests are demonstrated in Section~\ref{Numerical experiments}.
Conclusions are in Section~\ref{sec:conclusion}.
\section{Algorithm development}

\label{Algorithm development}

\subsection{Approach}

Let $l(\{Z_{t}\},\sum_{k\in K}|U_{t,k}|^{2})$ be a minus-log-likelihood of the
observed $\{Z_{t}\}$, $t\in T,$ and the complex-valued images at the sensor
plane are $U_{t,k}=A_{t,k}U_{o,k}$. The minus-log-likelihood is a fidelity term measuring  the misfit between the
observations and the prediction of the intensities of $U_{t,k}$ summarized
over the spectral interval.
Various  inverse imaging computational  methods have been developed  under variational
optimization formulation using one fidelity term and one prior term. 

We start from this tradition and introduce an unconstrained maximum likelihood optimization to reconstruct the
object 3D cube $\{U_{o,k}\},$ $k\in K,$ from the criterion of the form:
\begin{equation}
\min_{\{U_{o,k}\}}l(\{Z_{t}\},\sum_{k\in K}|A_{t,k}U_{o,k}|^{2})+f_{reg}%
(\{U_{o,k}\}), \label{first unconst opt}%
\end{equation}

where the second summand is an object prior.

The straightforward optimization in \eqref{first unconst opt} is too complex. The alternating direction method of
multipliers (ADMM) provides a valuable alternative to this sort of problems \cite{ADMM_1},\cite{ADMM_2}, \cite{ADMM_3}, \cite{ADMM_4}. The following logic
leads from \eqref{first unconst opt} to ADMM.

Let us reformulate \eqref{first unconst opt} as a constrained optimization:%
\begin{align}
&  \min_{\{U_{t,k},\text{ }U_{o,k}\}}l(\{Z_{t}\},\sum_{k\in K}|U_{t,k}%
|^{2})+f_{reg}(\{U_{o,k}\}),\label{constr-opt}\\
&  \text{subject to }U_{t,k}=A_{t,k}U_{o,k}\text{.}\nonumber
\end{align}

The problems \eqref{first unconst opt} and \eqref{constr-opt} are equivalent
with an advantage  introduced by $U_{t,k}$ which can be treated as a
splitting variable such that optimization can be arranged as sequential on
$U_{t,k}$ and $U_{o,k}$ of the two loss functions $l()$ and $f_{reg}()$. \ The
concept of splitting variables are at the root of many modern optimization methods (e.g. \cite{ADMM_2}, \cite{ADMM_3}).

One of the popular ideas to resolve \eqref{constr-opt} is
to replace it by the unconstrained formulation with  the  parameter $\gamma>0:$
\begin{align}
J  &  =l(Z_{t},\sum_{k}^{{}}|U_{t,k}|^{2})+\frac{1}{\gamma}\sum_{k,t}^{{}%
}||U_{t,k}-A_{t,k}U_{o,k}||_{2}^{2}+  f_{reg}(\{U_{o,k}\})
\label{likelihood}
\end{align}

The second summand is the quadratic penalty for the difference between
$A_{t,k}U_{o,k}$ and the splitting $U_{t,k}$. In optimization of
\eqref{likelihood}, $U_{t,k}$ $\rightarrow A_{t,k}U_{o,k}$ as $\gamma
\rightarrow0$. 
The minimization algorithm iterates $\min_{\{U_{k,t}\}}J$,  provided
given $U_{o,k}$\textcolor{black}{,} and $\min_{\{U_{o,k}\}}J$ provided fixed $U_{t,k}$:%
\begin{align}
U_{t,k}^{(s+1)}  &  =\arg\min_{U_{t,k}}(l(Z_{t},\sum_{k}^{{}}|U_{t,k}%
|^{2})+\frac{1}{\gamma}\sum_{k,t}^{{}}||U_{t,k}-A_{t,k}U_{o,k}^{(s)}||_{2}%
^{2}),\label{Algoritm0}\\
U_{o,k}^{(s+1)}  &  =\arg\min_{U_{o,k}}(f_{reg}(\{U_{o,k}\})+\frac{1}{\gamma
}\sum_{k,t}^{{}}||U_{t,k}^{(s+1)}-A_{t,k}U_{o,k}||_{2}^{2}).\nonumber
\end{align}

It is recommended in these iterations to take $\gamma$ varying and going to
smaller values $\gamma\rightarrow0$. The performance of the algorithm depends
on this parameter.

A valuable counterpart to \eqref{likelihood} with  a decreasing 
sensitivity to  $\gamma$ is a reformulation of \eqref{constr-opt}
with the Lagrange multipliers $\Lambda_{t,k}$ and the augmented Lagrange loss
function. For the complex-valued variables this counterpart is of the form \cite{Complex-ADNN}: %
\begin{align}
J(\{U_{t,k},U_{o,k},\Lambda_{k,t}\})=l(Z_{t},\sum_{k}^{{}}|U_{t,k}|^{2}%
)+\frac{1}{\gamma}\sum_{k,t}^{{}}||U_{t,k}-A_{t,k}U_{o,k}||_{2}^{2}+  &
\label{Lagrange formulation}\\
f_{reg}(\{U_{o,k},k\in K\})+2\frac{1}{\gamma}\operatorname{Re}(\sum_{k,t}^{{}}\Lambda
_{t,k}^{H}(A_{t,k}U_{o,k}-U_{k,t}))  &  .\nonumber
\end{align}
Here the Lagrange multipliers are complex-valued, $\Lambda_{t,k}\in
\mathbb{C}^{M}$, and the subscript '$^{H}$' stays for the Hermitian
transpose. The complex-valued variables introduce specific features to this
formulation, differing it from the conventional  real-valued one. 

The alternating direction algorithm of multipliers (ADMM) for the Lagrangian
\eqref{Lagrange formulation} is composed of iterations \cite{Complex-ADNN}:
\begin{align}
U_{t,k}^{(s+1)}  &  =\arg\min_{U_{t,k}}J(\{U_{t,k},U_{o,k}^{(s)},\Lambda
_{t,k}^{(s)}\}),\label{Iterations0}\\
U_{o,k}^{(s+1)}  &  =\arg\min_{U_{o,k}}J(\{U_{t,k}^{(s+1)},U_{o,k},\Lambda
_{t,k}^{(s)}\}),\nonumber\\
\Lambda_{t,k}^{(s+1)}  &  =\Lambda_{t,k}^{(s)}-\frac{1}{\gamma} (U_{t,k}^{(s+1)}-A_{t,k}%
U_{o,k}^{(s+1)}).\nonumber
\end{align}
The last equation updates the Lagrange multipliers.

It can be verified that
\begin{align*}
||U_{t,k}-A_{t,k}U_{o,k}-\Lambda_{t,k}||_{2}^{2}=||U_{t,k}%
-A_{t,k}U_{o,k}||_{2}^{2}+ 
2\operatorname{real}(\Lambda_{t,k}^{H}(A_{t,k}U_{o,k}-U_{t,k}))+||\Lambda
_{k,t}^{H}||_{2}^{2}. 
\end{align*}

As $||\Lambda_{t,k}^{H}||_{2}^{2}$ does not depend  on $U_{t,k}$ and $U_{o,k}%
$, the iterations \eqref{Iterations0} can be rewritten for
\eqref{Lagrange formulation} as%
\begin{align}
U_{t,k}^{(s+1)}  &  =\arg\min_{U_{t,k}}(l(Z_{t},\sum_{k}^{{}}|U_{t,k}%
|^{2})+\frac{1}{\gamma}\sum_{k,t}^{{}}||U_{t,k}-A_{t,k}U_{o,k}^{(s)}%
-\Lambda_{t,k}^{(s)}||_{2}^{2}),\label{ADMM}\\
U_{o,k}^{(s+1)}  &  =\arg\min_{U_{o,k}}(f_{reg}(\{U_{o,k},k\in K\}+\frac
{1}{\gamma}\sum_{k,t}^{{}}||U_{t,k}^{(s+1)}-A_{t,k}U_{o,k}-\Lambda_{t,k}%
^{(s)}||_{2}^{2}),\nonumber\\
\Lambda_{t,k}^{(s+1)}  &  =\Lambda_{t,k}^{(s)}-(U_{t,k}^{(s)}-A_{t,k}%
U_{o,k}^{(s)}),\nonumber
\end{align}
where the Lagrange multipliers $\Lambda_{t,k}^{(s)}$ are replaced by the scaled ones defined as $\gamma \Lambda_{t,k}^{(s)}$.

We exploit the ADMM algorithm in this  form for the consider HS phase
retrieval problem. In the following sections, we derive the solutions for the optimizations in \eqref{ADMM}.

\subsection{Minimization on $U_{t,k}$}

\subsubsection{Gaussian observations\label{Gaussian}}

the loss function in the first row of
\eqref{ADMM} is of the form%
\begin{equation}
J=\frac{1}{\sigma^{2}}\sum_{t}^{{}}||Z_{t}-\sum_{k}^{{}}|U_{t,k}|^{2}%
||_{2}^{2}+\frac{1}{\gamma}\sum_{k,t}^{{}}||U_{t,k}-A_{t,k}U_{o,k}%
-\Lambda_{k,t}||_{2}^{2}. \label{Gauss_criterion}%
\end{equation}

For minimization $\min_{U_{t,l}}J$, we calculate the derivatives $\partial
J/\partial U_{t,l}^{\ast}$ and consider the necessary minimum conditions
$\partial J/\partial U_{t,l}^{\ast}(r)=0$. These
calculations lead to a set of complex-valued cubic
equations with respect to $U_{t,l}(r)$:%
\begin{equation}
\lbrack\frac{2}{\sigma^{2}}(\sum_{k\in K}|U_{t,k}(r)|^{2}-Z_{t}(r))+\frac
{1}{\gamma}]U_{t,l}(r)=\frac{1}{\gamma}(A_{t,l}U_{o,l}(r)+\Lambda
_{t,l}(r)),\text{ }l\in K\text{.} \label{Ullt}%
\end{equation}
Here, $r$ stays for the spacial coordinates instead of $(x,y)$ in $2D$ image representations. Note that these
equations are separated on $r$ as well as with respect to $t.$

The following manipulations resolve the set $\eqref{Ullt}$. Calculating the squared
absolute values for both sides of these equations and producing summation
on $l\in K$ , we arrive to%
\begin{equation}
\lbrack\frac{2}{\sigma^{2}}\sum_{k\in K}|U_{t,k}(r)|^{2}-Z_{t}(r)+\frac
{1}{\gamma}]^{2}\sum_{k\in K}|U_{t,k}(r)|^{2}     =\label{original cubick} 
\sum_{k\in K}|\frac{1}{\gamma}(A_{t,k}U_{o,k}(r)+\Lambda_{t,k}(r))|^{2}.   
\end{equation}

With the notations:
\begin{equation}
x\triangleq\sum_{k\in K}|U_{t,k}(r)|^{2};\text{ }q\triangleq\sum_{k\in K}%
|(A_{t,k}U_{o,k}(r)+\Lambda_{t,k}(r))|^{2}, \label{Notations}%
\end{equation}
we rewrite \eqref{original cubick} in a compact form as a cubic Cardano's
equation with respect to $x$:%

\begin{align}
&  ax^{3}+bx^{2}+cx+d=0,\label{CARDAN_EQ}\\
&  a=(\frac{2\gamma}{\sigma^{2}})^{2};\text{ }b=\frac{4\gamma}{\sigma^{2}%
}(1-\frac{2\gamma}{\sigma^{2}}Z_{t}(r));\text{ }c=(1-\frac{2\gamma}{\sigma
^{2}}Z_{t}(r));\text{ }\nonumber\\
&  d=-q.\nonumber
\end{align}

The Cardano's formulas give the solutions for \eqref{CARDAN_EQ}. The
coefficients of this equation are real-valued. It has three roots, which,
depending on the discriminant $D$, are real-valued for $D\leq0$ and 
one root is real and two others are complex-valued for $D>0$.

We are looking for real-valued nonnegative solutions $x\geq0$. If there are several such solutions, we select the one which provides a smallest value to the
criterion \eqref{Gauss_criterion}.

If such $\hat{x}$ is selected\textcolor{red}{,} the solution for $U_{k,t}(r)$ according to
\eqref{Ullt} is as follows:
\begin{equation}
\hat{U}_{t,k}(r)=(A_{t,k}U_{o,k}(r)+\Lambda_{t,k}(r))/(1+\frac{2\gamma}%
{\sigma^{2}}(\hat{x}_{t}(r)-Z_{t}(r))). \label{B_lt}%
\end{equation}

Equation \eqref{CARDAN_EQ} should be solved for each $r$ and $t$ separately.  It follows that the criterion for selection of the best
$\hat{x}_{t}(r)$ and  $\hat{U}_{t,k}(r)$ should be pixel-wise:%
\begin{equation}
J_{t}(r)=\frac{1}{\sigma^{2}}(Z_{t}(r)-\sum_{k\in K}^{{}}|\hat{U}%
_{t,k}(r)|^{2})^{2}+\frac{1}{\gamma}\sum_{k\in K}^{{}}|\hat{U}_{t,k}%
(r)-A_{t,k}U_{o,k}(r)-\Lambda_{t,k}(r)|^{2}, \label{localCriterion}%
\end{equation}
where $\hat{U}_{t,k}(r)$ is given by \eqref{B_lt}.

Thus, the solution for $\hat{U}_{t,k}(r)$ is obtained by the following three
stage procedure: (1) Solution of \eqref{CARDAN_EQ}; (2) Analysis of the roots
of the Cardano's equation selecting the one minimizing \eqref{localCriterion}; (3)
Calculation of $\hat{U}_{t,k}(r)$ by \eqref{B_lt}.\\ In our MATLAB implementation of the algorithm  the calculations for all pixels $r$ are done  in parallel.

The solution $\hat{U}_{t,k}(r)$ is a crucial point of the developed algorithm as
it allows to extract spectral information from the total observations summarized over
entire spectral components. This solution can be treated as a synchronous
detector with a modulating signal $(A_{t,k}U_{o,k}(r)+\Lambda_{t,k}(r))$
applied to  the residual $(\hat{x}_{t}(r)-Z_{t}(r))$, where $\hat{x}_{t}(r)$ is an
estimate of $Z_{t}(r)$.

The spectral resolution of this detector is restricted by differences in the spectral
properties of $A_{t,k}U_{o,k}(r)$, i.e. in the spectral properties of the
image formation operator $A_{t,k}$ and in the spectral properties of the object
$U_{o,k}(r)$.
\bigskip
\subsubsection{Poissonian observations\label{Poissonian}}

The observations take random integer values which can be interpreted as 
a counted number of photons detected by the sensor. This discrete distribution
has a single parameter $\mu$ and is defined by the formula $p(Z_{t}%
(r)=n)=\exp(-\mu)\dfrac{\mu^{n}}{n!}$, where $Z_{t}(r)$ is $r-th$ pixel of the
intensity vector $Z_{t}$. Here $p(Z_{t}(r)=n)$ is the probability that a
random $Z_{t}(r)$ takes value $n$, where $n\geq0$ is an integer. The parameter
$\mu$ is the intensity flow of Poissonian random events. The parameter $\mu$
is different for different experiment $t$ and $r$ with values $\mu=Y_{t}(r)$.
The probabilistic Poissonian observation model is given by the formula%
\begin{equation}
p(Z_{t}(r)=n)=\exp(-Y_{t}(r)\chi)\frac{(Y_{t}(r)\chi)^{n}}{n!}\text{,}
\label{observation_31}%
\end{equation}
where $\chi>0$ is introduced as a scaling parameter for photon flow.

According to the properties of the Poisson distribution, we have for the
mean value and the variance of the observed $Z_{t}(r)$, $E\{Z_{t}%
(r)\}=var\{Z_{t}(r)\}=Y_{t}(r)\chi$. In these formulas, $\chi$ defines the level
of the random component in the signal and can be interpreted as an exposure
time. A larger $\chi$ means a larger exposure time, and a larger number of the
photons with the intensity $Y_{t}(r)\chi$ is recorded. The signal-to-noise ratio  $E\{Z_{t}%
(r)\}^{2}/var\{Z_{t}(r)\}=Y_{t}(r)\chi$ takes larger values for larger $\chi$.

The minus-log-likelihood for \eqref{observation_31} gives $Y_{t}(r)\chi
-Z_{t}(r)\log(Y_{t}(r)\chi)$ and the criterion \eqref{likelihood} takes the
form
\begin{align}
J  &  =\sum_{t,r}^{{}}(\chi\sum_{k}^{{}}|U_{t,k}(r)|^{2}-Z_{t}(r)\log(\sum
_{k}^{{}}|U_{t,k}(r)|^{2}\chi))+\label{Poiss_criterion}\\
&  \frac{1}{\gamma}\sum_{k,t}^{{}}||U_{t,k}-A_{t,k}U_{o}-\Lambda_{t,k}%
||_{2}^{2}+f_{reg}(\{U_{o,k}\}_{k\in K}).\nonumber
\end{align}

The necessary minimum condition on $U_{t,k}$ can be calculated as $\partial
J/\partial U_{l,t}^{\ast}(r)=0$ leading to the set of the nonlinear
equations with respect to $U_{t,l}(r):$%
\begin{equation}
(\chi-Z_{t}(r)/\sum_{k\in K}^{{}}|U_{t,k}(r)|^{2}+\frac{1}{\gamma}%
)U_{t,l}(r)=\frac{1}{\gamma}(A_{t,l}U_{o,l}(r)+\Lambda_{t,l}(r)).
\label{PoissEq}%
\end{equation}
To resolve the problem, we use the methodology applied for the  Gaussian data. 
We calculate the sums over
$l$ for the squared magnitude values of the left and right sides of
\eqref{PoissEq} and using the notations \eqref{Notations} obtain the set of the 
quadratic equations for $x$ :%
\begin{align}
&  ax^{2}+bx+c=0,\\
& a=(1+\gamma\chi)^{2};\text{ }b=-2(1+\gamma\chi)\gamma Z_{t}(r)-q;\text{
}\nonumber\\
&  c=(\gamma Z_{t}(r))^{2};\text{ }q=\sum_{k\in K}|(A_{t,k}U_{o,k}%
(r)+\Lambda_{t,k}(r))|^{2}.
\label{Quadratic}%
\end{align}

As $b^{2}-4ac=q^{2}+4(1+\gamma\chi)\gamma Z_{t}(r)\geq0$, the quadratic
equations have two real roots. The estimate for $U_{k,t}(r)$ corresponding to
the solution $\hat{x}$ according to \eqref{PoissEq} is as follows
\begin{equation}
\hat{U}_{t,k}(r)=(A_{t,k}U_{o,k}(r)+\Lambda_{t,k}(r))/(1+\gamma\chi+\frac
{1}{\hat{x}_{t}(r)}\gamma Z_{t}(r))\text{, }k\in K. \label{Poisson_Solution}
\end{equation}

As there are two real-valued roots, to select the proper root which is
nonnegative and the best minimizer for $J$, we use pixelated summands of this
criterion with fixed $t$ and $r:$
\begin{equation}
J_{t}(r)   =(\chi\sum_{k\in K}^{{}}|\hat{U}_{t,k}(r)|^{2}-Z_{t}(r)\log
(\sum_{k\in K}^{{}}|\hat{U}_{t,k}(r)|^{2}\chi))\label{PoissAnalysis} 
  +\frac{1}{\gamma}\sum_{k\in K}^{{}}|\hat{U}_{t,k}(r)-A_{t,k}U_{o,k}%
(r)-\Lambda_{t,k}(r)|^{2}.
\end{equation}
\ 

Both solutions for $\min_{U_{k,t}}J$ (Gaussian and Poissonian) can be
interpreted as the proximity operators being obtained minimizing the
likelihood items (first item in \eqref{likelihood}) regularized by the
quadratic penalty (second item in\ \eqref{likelihood}). With the standard
compact notation for the proximity operators \cite{ProximalOperator-2019}, \cite{ProximalOperators},
 we denote these solutions as:%
\begin{equation}
\hat{U}_{t,k}=prox_{f\gamma}(A_{t,k}U_{o,k}+\Lambda_{k,t})\text{,}
\label{proximity}%
\end{equation}
where $f$ stays for the minus-log-likelihood part of $J$ and $\gamma>0$ is a parameter.

These proximity operators resolve two problems. Firstly, the complex domain
spectral components $U_{t,k}(r)$ are extracted from the observations where all of
them are measured jointly as the total power of the signal. Thus, we obtained
the spectral analysis of the observed signals. Secondly, the noisy
observations are filtered with the power controlled by the parameter $\gamma$
compromising the noisy intensity observations $Z_{t}$ and the power of the
predicted signal $A_{t,k}U_{o,k}$ at the sensor plane.
We introduce a term
'Spectral  Proximity  Operator'  (SPO) for \eqref{proximity}.  

The proximity operators have been already used for phase retrieval in its
monochromatic version (single wavelength, $k=1$) in  \cite{Unser},
\cite{VK_Gauss_POiss}. The cubic and quadratic
equations have appeared for Gaussian and Poissonian observations,
respectively, but they are scalar equations, while for the considered
HS problem we need resolve the sets of cubic and quadratic equations. The
spectral resolution properties of these new HS operators define
their unique peculiarity.

\subsection{Minimization on $U_{o,k}$}

Minimizing \eqref{ADMM} on $U_{o,k}$ we drop the regularization term
$f_{reg}(\{U_{o,k},k\in K\})$ and replace its effects by joint filtering of
$\{U_{o,k},k\in K\}$. This approach is in line with a recent tendency to use
efficient filters for regularization of ill-posed solutions without
formalization of the regularization penalty in variational setups of
imaging problems (e.g. \cite{Baraniuk}). As a replacement for the prior we use the Complex Cube Filter
(CCF) \cite{shevkunov_CCF_2020}. 

Then, the solution for $U_{o,k}$ is of the form
\begin{equation}
U_{o,k}=\frac{\sum_{t}A_{t,k}^{H}(U_{k,t}-\Lambda_{k,t})}{\sum_{t}%
A_{t,k}^{H}A_{t,k}+reg}, \label{Object}%
\end{equation}
where the regularization parameter $reg>0$ is included if $\sum_{t}%
A_{t,k}^H A_{t,k}$  is singular or ill-conditioned.

\subsection{ HS Complex Domain Phase Retrieval (HSPhR) algorithm}
\label{sec:algorithm}
Let us present the iterations of the algorithm developed based on the above
solutions.  
\bigskip\\
Input data $\{Z_{t},t\in T\},$ operators $\{A_{t,k},k\in K,$ $t\in T\}.$

(1) Initialization: $\{U_{o,k}^{(0)}$, $\Lambda_{t,k}^{(0)}=0,$ $k\in K,$
$t\in T\};$

Repeat for $s=1,2,...,$ $n$,

(2) Forward propagation: $U_{t,k}^{(s)}=A_{t,k}U_{o,k}^{(s)}$, $k\in K,$ $t\in
T;$

(3) Update $U_{k,t}^{(s)}$ by SPOs: $\hat{U}_{t,k}%
^{(s)}=prox_{f\gamma}(U_{t,k}^{(s)}+\Lambda_{t,k}^{(s)})$, $k\in K,$ $t\in T;$

(4) Update Lagrange variables: $\Lambda_{t,k}^{(s)}=\Lambda_{t,k}^{(s)}%
-(\hat{U}_{t,k}^{(s)}-U_{t,k}^{(s)});$

(5) Backward propagation and preliminary object estimation:
\[
U_{o,k}^{(s)}=\frac{\sum_{t}A_{t,k}^{H}(\hat{U}_{t,k}^{(s)}-\Lambda
_{t,k}^{(s)})}{\sum_{t}A_{t,k}^{H}A_{t,k}+reg},k\in K;
\]

(6) Update of $U_{o,k}^{(s)}$ by complex domain filtering:
\[
U_{o,k}^{(s+1)}=(1-\beta_{s})U_{o,k}^{(s)}+\beta_{s} CCF(\{U_{o,k}^{(s)},k\in K\}),k\in K;
\]

(7)\ Return $U_{o,k}^{(n+1)}$.
\bigskip\\
The complex domain initialization (Step 1) is required for the considered spectral domain $k\in K$. In our experiments, we assume 2D random white-noise Gaussian distribution for phase and a uniform 2D positive distribution on (0,1] for amplitude, which are independent for each $k$. The Lagrange multipliers are initialized by zero values, $\Lambda_{k,t}=0$.  The forward propagation is produced for all $k\in K
$ and $t\in T$ (Step 2). The update of the wavefront at the sensor plane (Step
3) is produced by the proximal operators. For the Gaussian observations, this
operator is defined by  \eqref{CARDAN_EQ}-\eqref{B_lt} and for the
Poissonian observations by  \eqref{Quadratic}-\eqref{Poisson_Solution}.
It requires to solve the polynomial equations,  cubic or quadratic for the Gaussian and
Poissonian case, respectively. In Step 4, the Lagrange variables are updated. The backward propagation of the wavefront from the sensor plane to the object plane is combined with an update of the spectral object estimate at Step 5. The sparsity based regularization by Complex Cube Filter
(CCF) is relaxed by the weight-parameter $0<\beta<1$ at Step 6. The iteration number is fixed (Step 6) in this algorithm implementation.

The CCF algorithm is designed to deal with 3D complex-valued cube data and introduced in details in \cite{shevkunov_CCF_2020}. Examples of its application and features of this
algorithm can be seen in \cite{CCF-ICIP2020}, \cite{CCF-SENSOR}.

A few notes on this algorithm. The CCF algorithm is based on SVD analysis of
the HS complex-valued cube. It identifies an optimal subspace for the HS image
representation including both the dimension of the eigenspace and eigenimages
in this space. The Complex-Domain Block-Matching 3D (CDBM3D) algorithm
\cite{Katkovnik-2017-CDBM3D}, \cite{katkovnik2017complex} filters these eigenimages. Going from the
eigenimage space back to the original image space, we obtain the reconstruction
of the object cube. The CDBM3D is developed as an extension to the complex domain of the popular BM3D algorithm \cite{Dabov2007}. 
 CDBM3D is implemented in MATLAB as Complex
Domain Image Denoising (CDID) Toolbox \cite{katkovnik2017complex}. 
A sparsity modeling in CDBM3D is based on patch-wise 3D/4D  Block-Matching grouping and  3D/4D High-Order
Singular Decomposition (HOSVD) exploited for block-wise spectrum design, analysis and
filtering (thresholding and Wiener filtering). Three types of sparsity are developed in CDID corresponding to the following representations of complex-valued variables: I. Complex variables (3D grouping); II. Real and imaginary parts of complex variables (4D grouping); III. Phase and amplitude (4D grouping). In the forthcoming simulation tests we use the joint  real/imaginary sparsity (Type II). 
\section{Numerical experiments}

\label{Numerical experiments}
We present the  results of numerical experiments to evaluate the empirical
performance of the developed algorithms. 
Phase and amplitude imaging by the lensless computational microscopy system is demonstrated. The optical setup
illustrating the image formation is shown in Fig.\ref{fig:scheme}.
A broadband coherent laser beam, wavelength range $\Lambda=[400:700]$~nm, of the uniform intensity distribution impinges on a transparent thin object $O$. \textcolor{black}{The phase coding modulation mask
$\mathcal{M}_{t}$ is attached to the object. This mask is implemented by the spatial light modulator (SLM), which is a programmable device allowing to change the mask $\mathcal{M}_{t}$ (phase coding) from experiment-to-experiment.} 
The laser beam propagates through the object, goes through the modulation phase mask $\mathcal{M}_{t}$, and freely propagates in air to the sensor. 
After that, the intensity of the beam is registered by the sensor as a coded broadband diffraction pattern.
\begin{figure}[hbp]
\centering
\includegraphics[width=0.3\textwidth]{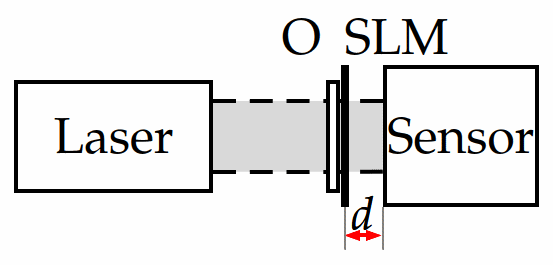}
\caption{Optical setup corresponding to the considered data formation model: 'Laser' is a broadband coherent light source, 'O' is an object, 'SLM' is a spatial light modulator ($\mathcal{M}_{t}$), and 'Sensor' is a registration camera.}
\label{fig:scheme}
\end{figure}

The free space propagation\ linking $U_{o,k}$ and $U_{t,k}$ is modeled by the
angular spectrum  operator calculated using the forward $\mathcal{F}\{\cdot\}$ and inverse $\mathcal{F}^{-1}\{\cdot\}$~Fourier
transforms:
\begin{equation}
U_{t,k}=\mathcal{F}^{-1}\{AS\circ\mathcal{F}\{\mathcal{M}_{t}\circ
U_{o,k}\}\}, \label{angularSpectrum}%
\end{equation}
where the angular spectrum transfer function in the $2D$ Fourier domain
$(f_{x},f_{y})$ is defined according to \cite{Goodman} as:
\begin{equation}
AS=\left\{
\begin{array}
[l]{l}%
\exp(jkd\sqrt{1-(\lambda f_{x})^{2}-(\lambda f_{y})^{2}})\text{, }(\lambda
f_{x})^{2}+(\lambda f_{y})^{2}\leq1\\
0\,\text{, otherwise}
\end{array}
\right.   \label{AS}%
\end{equation}
Here, $d$ is a distance from the object to the sensor plane equal to $2$~mm (see Fig.\ref{fig:scheme}).
In the discrete modeling, the sampling corresponds to the camera pixel size
equal to $3.45~\mu$m.

The equations \eqref{angularSpectrum} and \eqref{AS} are written for $2D$
variables, thus the coding phase mask $\mathcal{M}_{t}$ and the object $U_{o,k}$ are $2D
$ complex-valued functions.

The HS wavefronts and measurements are modeled by propagating the broadband
laser beam through the object and the phase modulation mask. Both of them are modeled by the
complex-valued transfer function \cite{Goodman}:%
\begin{equation}
g=a\circ\exp(-jkh(n_{\lambda}-1)), \label{object}%
\end{equation}
where $k=2\pi/\lambda$, $h$ is a thickness of the object/mask, $n_{\lambda
}$ is a  refractive index depending on  $\lambda$. 

The phase delay of the wavefront propagating through $g$ is defined by the argument of the
exponential function in \eqref{object}. 
It is assumed in our tests that the amplitude $a$ and the thickness $h$ are
spatially varying on $(x,y)$ but invariant with respect to the spectral (wavelength) arguments $k$ and $\lambda$. Nevertheless, the model \eqref{object} shows that the properties of the object and the masks are spectrally varying.


\begin{figure}[tbp]
\begin{minipage}[h]{0.49\linewidth}
\center{\includegraphics[width=1\linewidth]{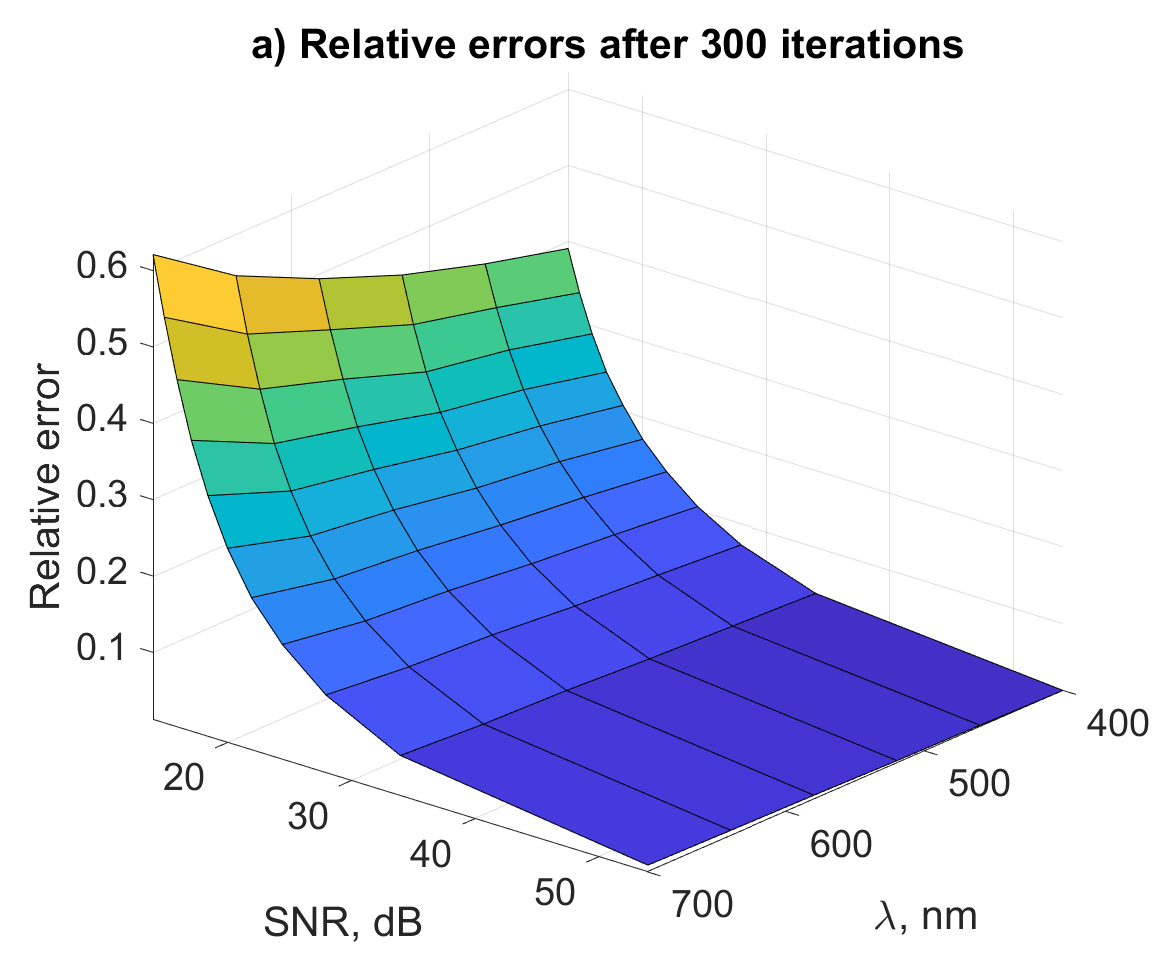} }
\end{minipage}
\hfill
\begin{minipage}[h]{0.49\linewidth}
\center{\includegraphics[width=1\linewidth]{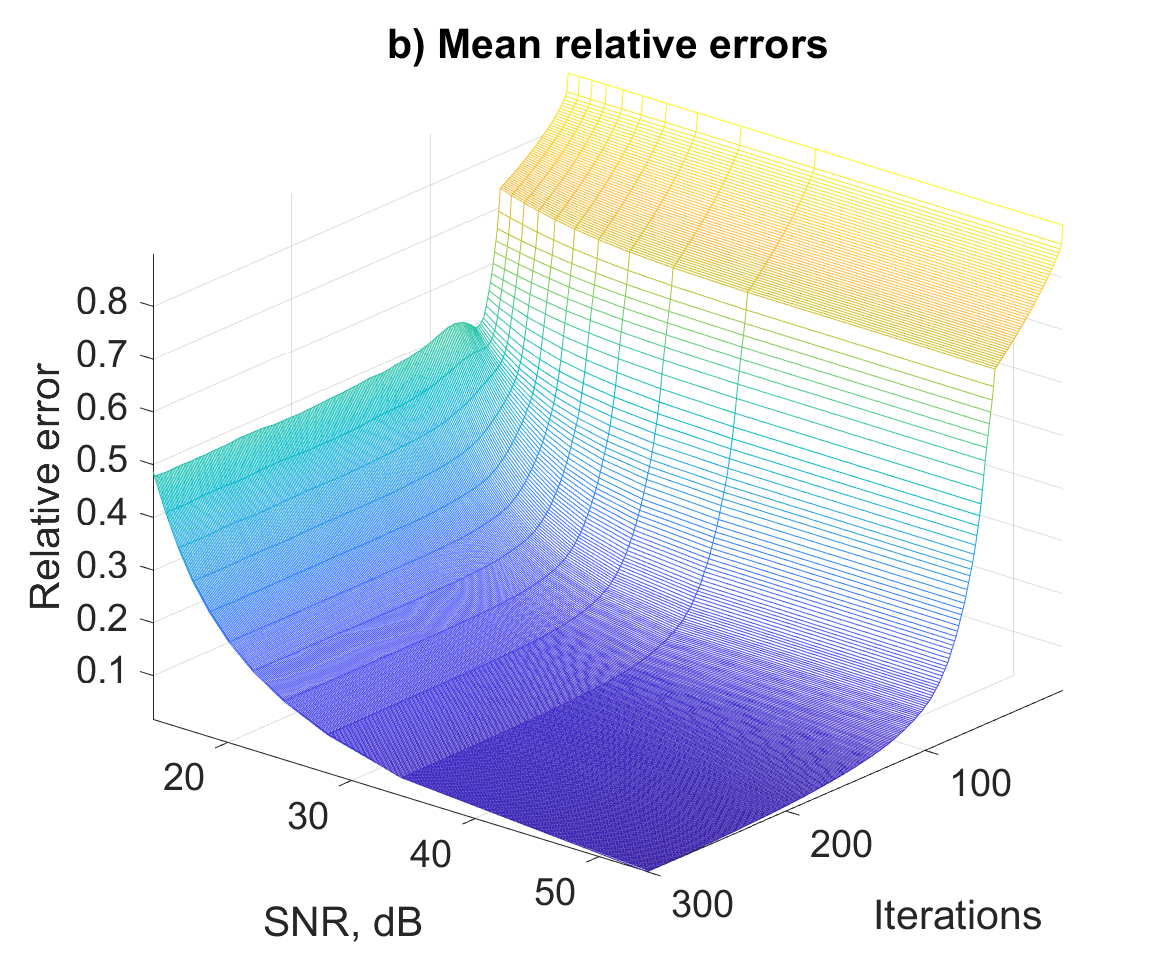} }
\end{minipage}
 \caption{Relative error maps for the HSPhR algorithm, Gaussian noise: a) $ERROR_{rel}$ as function of SNR and wavelength; b) $ERROR_{rel}$ as function of SNR and iteration number. In b) $ERROR_{rel}$ is averaged over the wavelengths. The iterations on CCF and the Lagrange variables are disabled up to 50th iteration.}
\label{fig:GaussErrors_CCFon_Don}
\end{figure}

\begin{figure}[hbp]
\begin{minipage}[h]{0.49\linewidth}
\center{\includegraphics[width=1\linewidth]{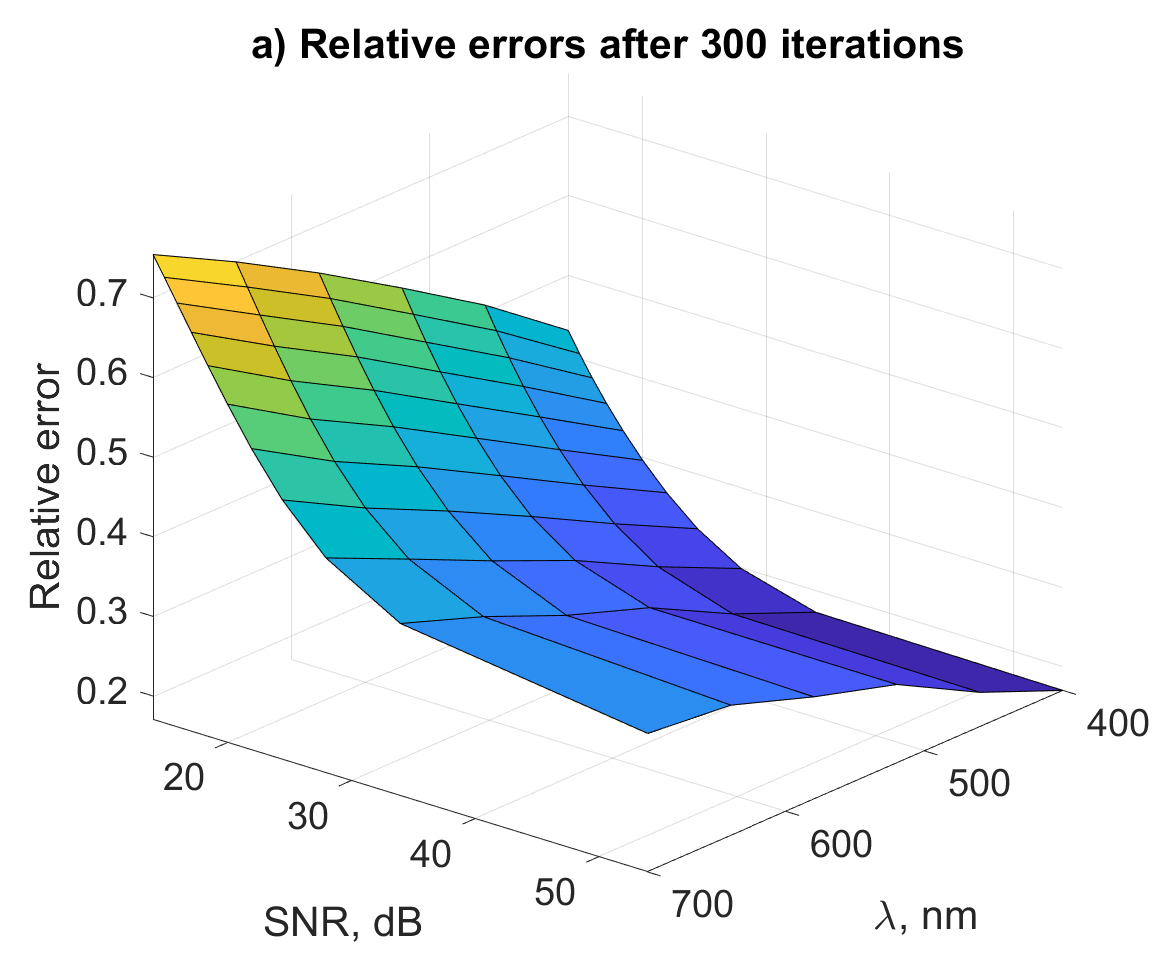} }
\end{minipage}
\hfill
\begin{minipage}[h]{0.49\linewidth}
\center{\includegraphics[width=1\linewidth]{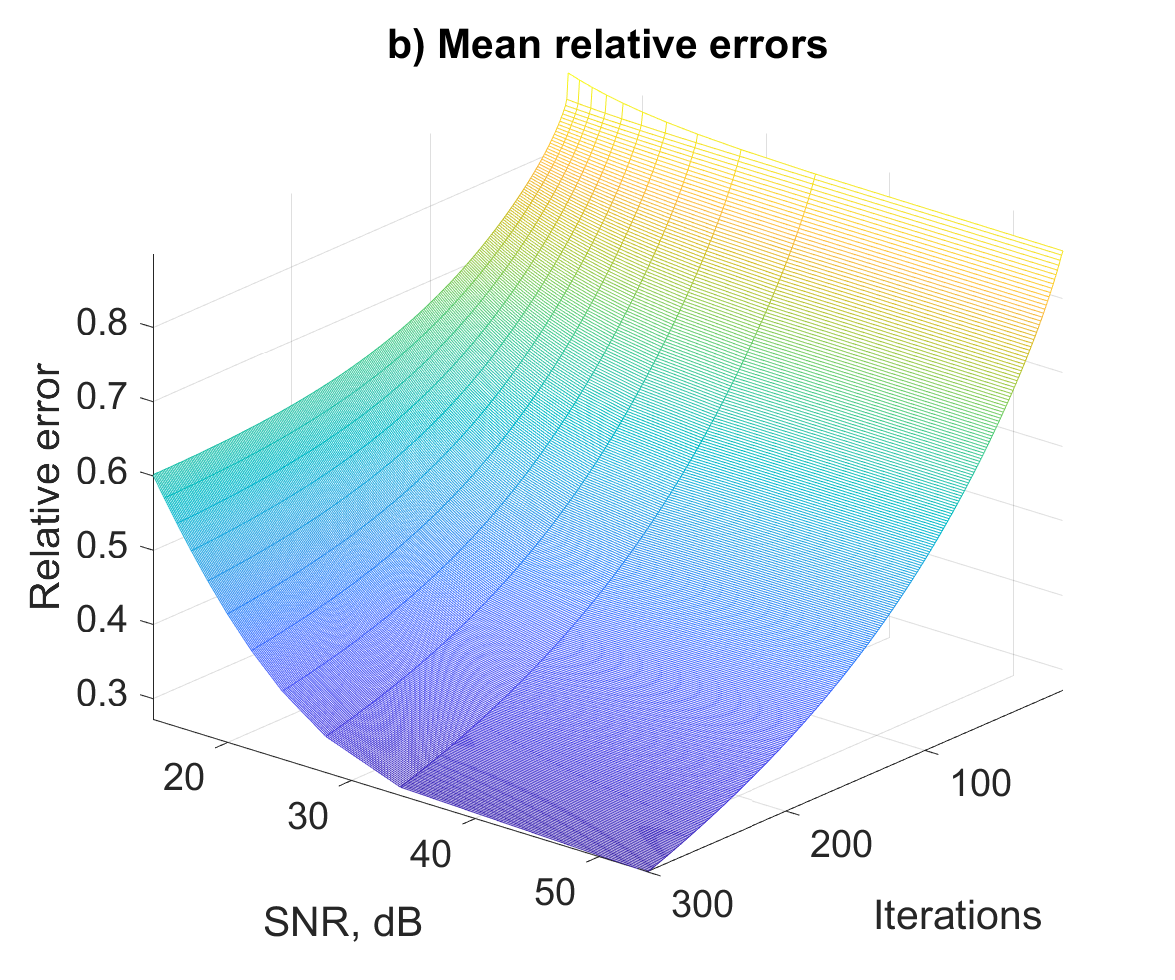} }
\end{minipage}
 \caption{Relative error maps for the HSPhR algorithm with disabled iterations on the Lagrange variables and the CCF filter, Gaussian noise: a) $ERROR_{rel}$ as function of SNR and wavelength; b) $ERROR_{rel}$ as function of SNR and iteration number. In b) $ERROR_{rel}$ is averaged over the wavelengths. The achieved accuracy is much worse than that for the HSPhR algorithm with enabled iterations on the Lagrange variables and the CCF filter shown in Fig.\ref{fig:GaussErrors_CCFon_Don}.
 } 
\label{fig:GaussErrors_CCFoff_Doff}
\end{figure}

\textcolor{black}{We assume that the amplitude and phase of the object are $64\times64$
images:  '\textit{peppers}' for amplitude \ and '\textit{cameraman}' for phase.}
We take these images quite different to test how far
phase and amplitude can be separated by the proposed algorithm. 
The phase is scaled in such way that the object phase delay for all $\lambda\in\Lambda$
would be in the range $[0$, $\pi]$. The amplitude $a$ is scaled to the
interval $(0,1]$.
For the coding mask, $a=1$, and $h$ is piece-wise invariant random with equal
probabilities taking one of  the following five values $[0,1,-1,1/2,-1/2]\cdot
\lambda_{\min}/4$, where $\lambda_{\min}=400$ nm. 

For each of the $T$ experiments, the
masks are generated independently. Thus, overall we have $T$ different masks.
The forward propagation (image formation) operator $A_{t,k}$ in \eqref{ourPR}
is defined by the propagation model \eqref{angularSpectrum} and the modulation
masks.
It is clear from the model \eqref{angularSpectrum} that the propagation
model is wavelength varying.
The same is true for the modulation mask, as it is fixed for each experiment, but its spectral properties vary with $\lambda$ according to \eqref{object}. 
As the phase mask is included in the operator
$A_{t,k}$, the complex-valued image at the sensor plane is defined as
$U_{t,k}=A_{t,k}U_{o,k}$, where the object is also spectrally varying 
according to \eqref{object}. The measured intensities are $Y_{t}=\sum_{k\in
K}|U_{t,k}|^{2}$.

In our simulation, we assume that the refractive index in \eqref{object} as a
function of $\lambda$ is known and calculated according to Cauchy's equation
\cite{born2013principles} with parameters taken for the glass BK7
\cite{Glass2014}.
The input laser beam is uniform in both phase and amplitude, the amplitude
is equal to $1$ and the phase is equal to $0$. We formulate the phase
retrieval as the reconstruction of $U_{o,k}$. 

The transfer function \eqref{object} is used for the calculation of the  observations $Y_{t}$ and the phase delay in the masks, and is not used in the algorithm iterations. 

We evaluate the accuracy of the complex-valued reconstruction by the relative
error criterion introduced in \cite{Candes2008}:%
\begin{equation}
ERROR_{rel}=\min_{\varphi\in\lbrack0,2\pi)}||\hat{x}\circ\exp(j\varphi
)-x||_{2}^{2}/||x||_{2}^{2}, \label{RelErrors}%
\end{equation}
where $x$ and $\hat{x}$ are the true signal and its estimate.



\textcolor{black}{ 
The efficiency of the developed algorithm is demonstrated in simulation tests with Gaussian and  Poissonian observations. 
In these experiments we present the results achieved by the developed algorithms using both the Lagrange multipliers (Step 4) and the $CCF$ filtering (Step 6) as it is in the HSPhR  algorithm , Subsection \ref{sec:algorithm}. In order to evaluate the value of these two key components of the algorithm, we show also results obtained when these components are disabled.}

\subsection{Accuracy as a function of $K$ and $N$.}
\textcolor{black}{ A dependence of the algorithm accuracy on the wavelength number $K$ and  the number of observations $T$ is of a special interest. In what follows, the noise level in observations is characterized by signal-to-noise ratio (SNR) in dB. We calculate  the accuracy criterion $ERROR_{rel}$ for
$K=[2,4,8,12]$ and $T=[2,6,12,24,36]$.
The wavelengths for the varying $K$ (spectral channels) are defined as uniformly covering the interval $[400, 700]$ nm. } The number of iterations is fixed to $n=300$.
}
The relative errors $ERROR_{rel}$ obtained in these experiments are shown in Table 1.

\begin{table}
\centering
\caption{$ERROR_{rel}$ versus number of experiments (masks) $T$ and  wavelengths $K$; $\lambda=400$~nm; number of iterations $n=300$; $SNR=54$~dB}
     \begin{tabular}{c||c|c|c|c|c}
        K\textbackslash T& 2 & 6 & 12 & 24 & 36   \\
         \hline\hline
        2 & 0.43 & 0.018 & 0.0062 & 0.0055 & 0.0061  \\
        4 & 0.68 & 0.097 & 0.011 & 0.0063 & 0.0074 \\
        8  & 0.82& 0.34 & 0.084 & 0.013 & 0.0103  \\
        12 & 0.85& 0.49 & 0.19 & 0.036 & 0.017  \\ 
         
    \end{tabular}
    
    \label{tab:my_label}
\end{table}

\textcolor{black}{ 
A larger value of $T$ for a fixed $K$ leads to more accurate reconstruction with smaller $ERROR_{rel}$.  We found that visually a good phase imaging is achieved provided $ERROR_{rel}<0.1$}. 
This threshold for $K=2$ is overcome for $T=6$. For $K=4,8,12$ it happens for $T=6,12,24$, respectively. Thus, we may conclude that  $T\geq 3K$ is sufficient for the accuracy sufficient for good phase imaging. The results in Table 1 are given for nearly noiseless  data with $SNR=54$ dB, but the conclusion that the inequality  $T\geq 3K$ is sufficient for visually good phase imaging holds for noisy data also.
This conclusion is one of the reasons to exploit in the forthcoming tests $T=18$ for $K=6$. 

\subsection{Gaussian observations \label{Gaussian_noise}}
The relative error maps for the HSPhR algorithm are shown in Fig.\ref{fig:GaussErrors_CCFon_Don} for SNR taking values  $[14:55]$~dB. The left image Fig.~\ref{fig:GaussErrors_CCFon_Don}(a) demonstrates the accuracy as a function of $SNR$ and $\lambda$, while the right image Fig.~\ref{fig:GaussErrors_CCFon_Don}(b) provides the accuracy as a function of $SNR$ and a number of iterations. In this latter image, the relative errors are averaged over $\lambda, K=6$. We may conclude from the left image that the acceptable quality imaging, $ERROR_{rel}<~0.1$, is achieved for all wavelengths provided that $SNR$ larger than 28~dB. The accuracy for smaller wavelengths is higher than the accuracy for larger values of wavelengths.

The convergence rate can be evaluated from the right image. After 200 iterations and for $SNR>28$ the accuracy becomes acceptable, $ERROR_{rel}<~0.1$.
The bend in the accuracy map in  Fig.~\ref{fig:GaussErrors_CCFon_Don}(b) well seen at iteration 50 is of special interest.  The algorithm's iterations on CCF and the Lagrange variables are disabled in this experiment up to this 50th iteration. Thus, we can observe the dramatic improvement in the convergence rate due to CCF and the Lagrange variables.  
We introduce this delay in the activation of CCF and the Lagrange variables only for demonstration of their efficiency. They can be activated from the first iterations that  improves the accuracy.

 In Fig.\ref{fig:GaussErrors_CCFoff_Doff} we show the relative error maps provided that both the Lagrange variables and $CCF$ are disabled in the  HS-CD-PhD algorithm completely. The degradation of the algorithm in values of $ERRORS_{rel}$ is clear in this case. In particular, it is demonstrated comparing the error maps in Fig.\ref{fig:GaussErrors_CCFon_Don} and Fig.\ref{fig:GaussErrors_CCFoff_Doff}. The relative errors in Fig.\ref{fig:GaussErrors_CCFoff_Doff} are always larger than 0.1, i.e., the accuracy of imaging is not acceptable.



\subsection{Poissonian observations \label{Poissonian_noise}}

\begin{figure}[tbp]
\begin{minipage}[h]{0.49\linewidth}
\center{\includegraphics[width=1\linewidth]{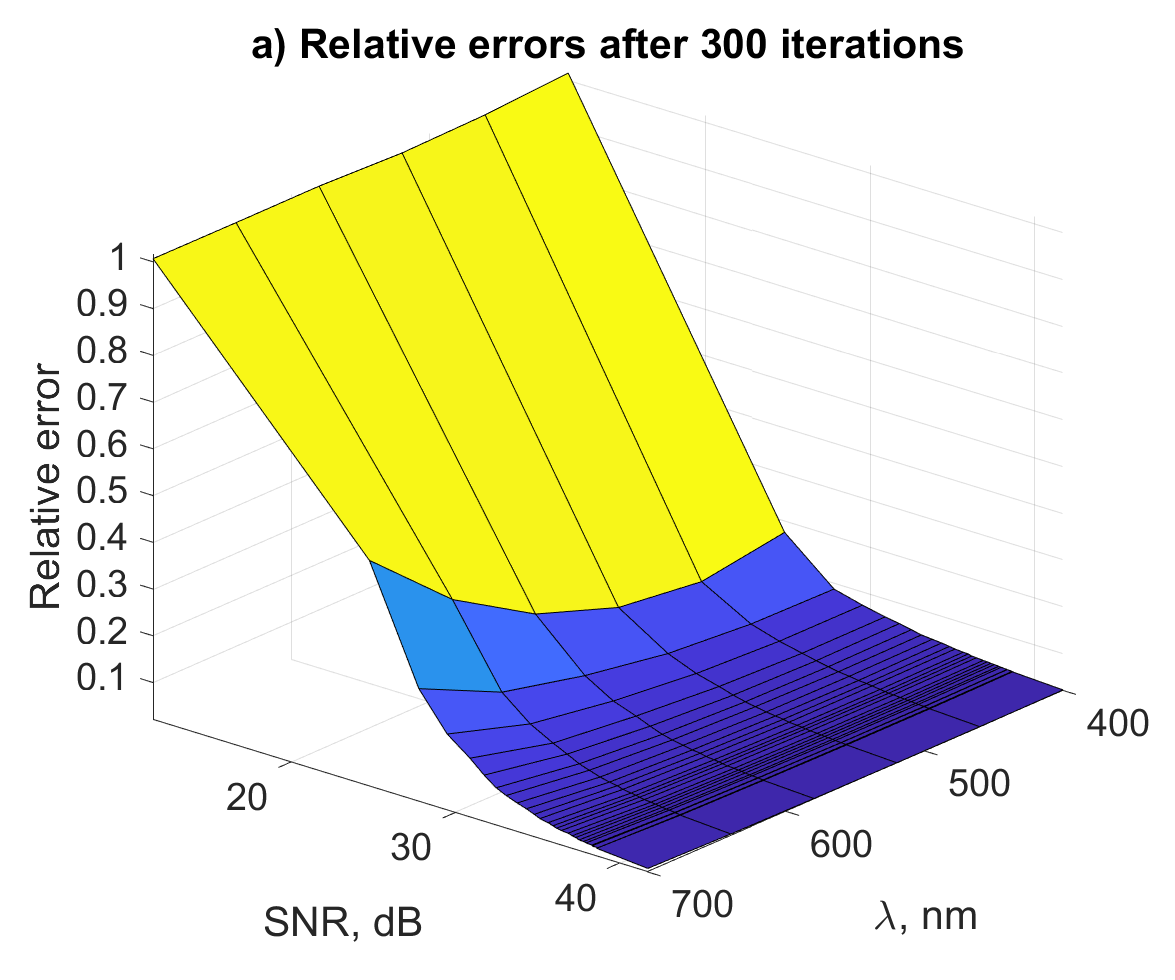} \\ a)}
\end{minipage}
\hfill
\begin{minipage}[h]{0.49\linewidth}
\center{\includegraphics[width=1\linewidth]{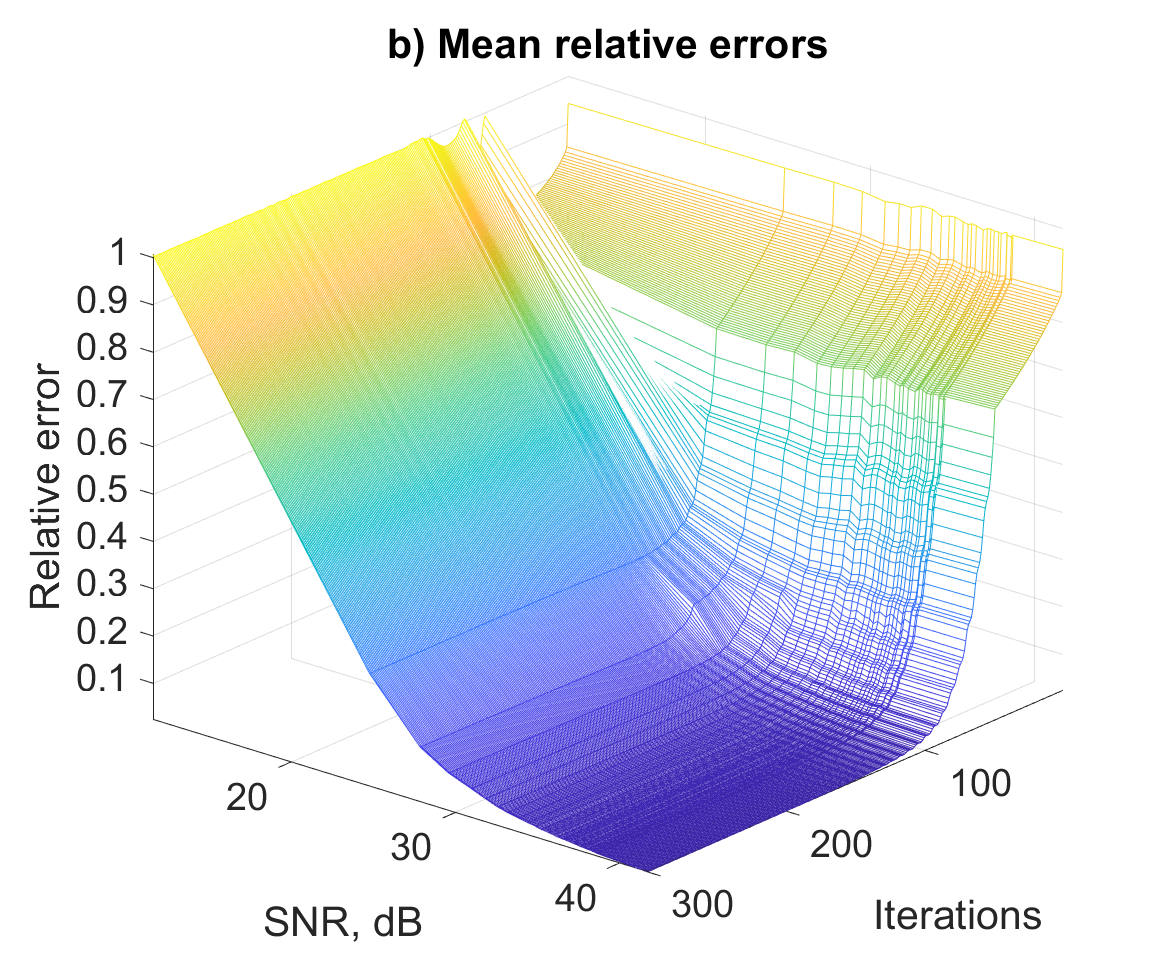} \\ b)}
\end{minipage}
 \caption{HSPhR reconstruction relative error maps for the object wavefronts in the case of Poisson noise depending on: a) SNR and wavelength after 300 iterations; b) SNR and iteration number. In b) error values are represented as mean value for all wavelengths. The iterations on CCF and the Lagrange variables are disabled up to 50th iteration.}
\label{fig:PoissonErrors_CCFon_Don}
\end{figure}

\textcolor{black}{For the scenario identical to considered for the case of Gaussian noise, we introduce results obtained for  Poissonian observations. The level of the noise is controlled by the   parameter $\chi$ which takes values corresponding SNR in the interval $[14:44]$~dB. 
The maps of the  relative errors are shown in Fig~\ref{fig:PoissonErrors_CCFon_Don}. 
The left image Fig.~\ref{fig:PoissonErrors_CCFon_Don}(a) demonstrates the accuracy as a function of $SNR$ and $\lambda$, while the right image Fig.~\ref{fig:PoissonErrors_CCFon_Don}(b) provides the accuracy as a function of $SNR$ and a number of iterations. In this latter image, the relative errors are averaged over $\lambda, K=6$. We may conclude from the left image that the acceptable quality imaging, $ERROR_{rel}<~0.1$, is achieved for all wavelengths provided that $SNR$ is larger than 32~dB. 
The convergence rate is well seen from the right image. After 150 iterations and for $SNR>32$~dB the accuracy becomes acceptable.
A visual bend in the accuracy map in  Fig.~\ref{fig:PoissonErrors_CCFon_Don}(b)  happened at the 50th iteration demonstrates the effects of  CCF and the Lagrange variables which are disabled up to the 50th iteration.  We can observe the dramatic improvement in the convergence rate due to CCF and the Lagrange variables. }

\begin{figure}[tbp]
\begin{minipage}[h]{0.49\linewidth}
\center{\includegraphics[width=1\linewidth]{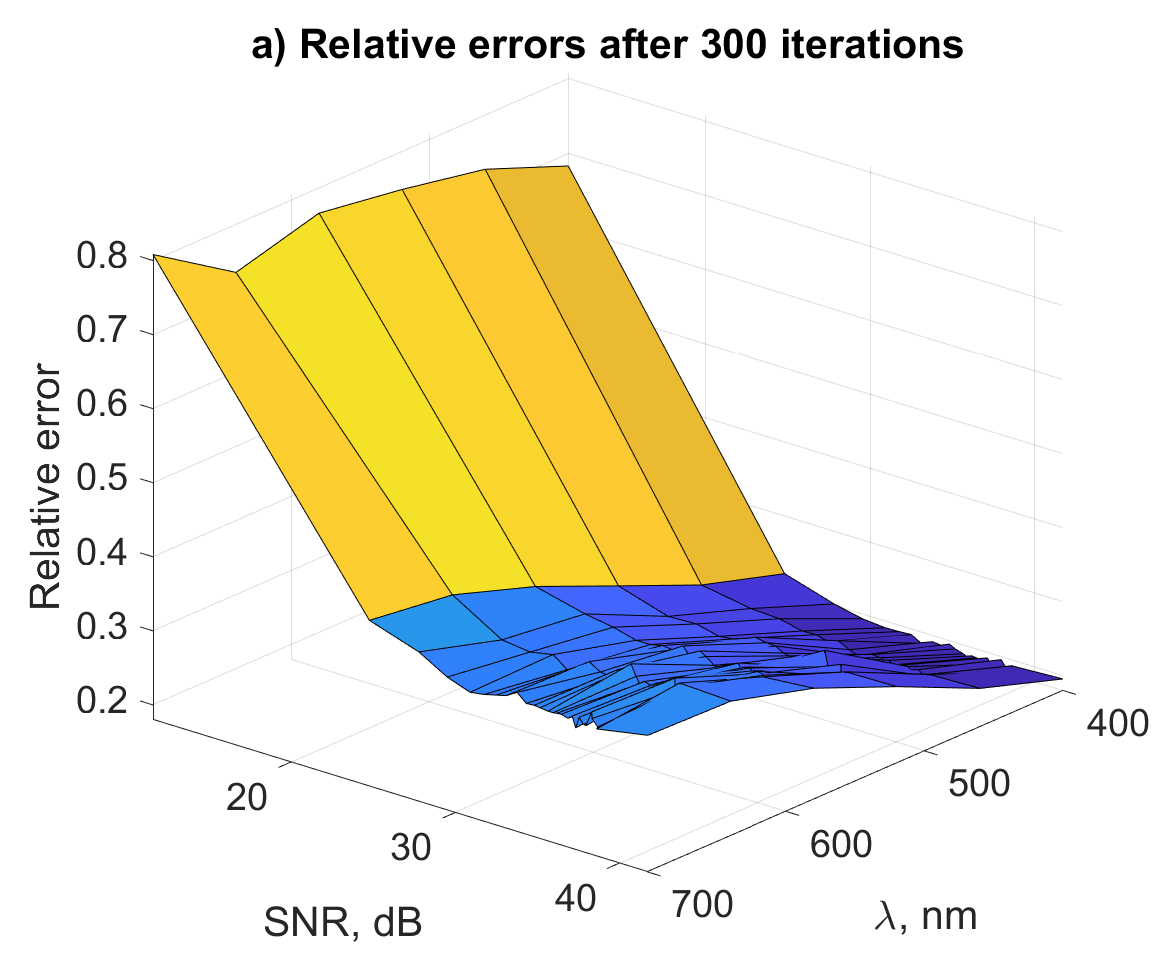} \\ a)}
\end{minipage}
\hfill
\begin{minipage}[h]{0.49\linewidth}
\center{\includegraphics[width=1\linewidth]{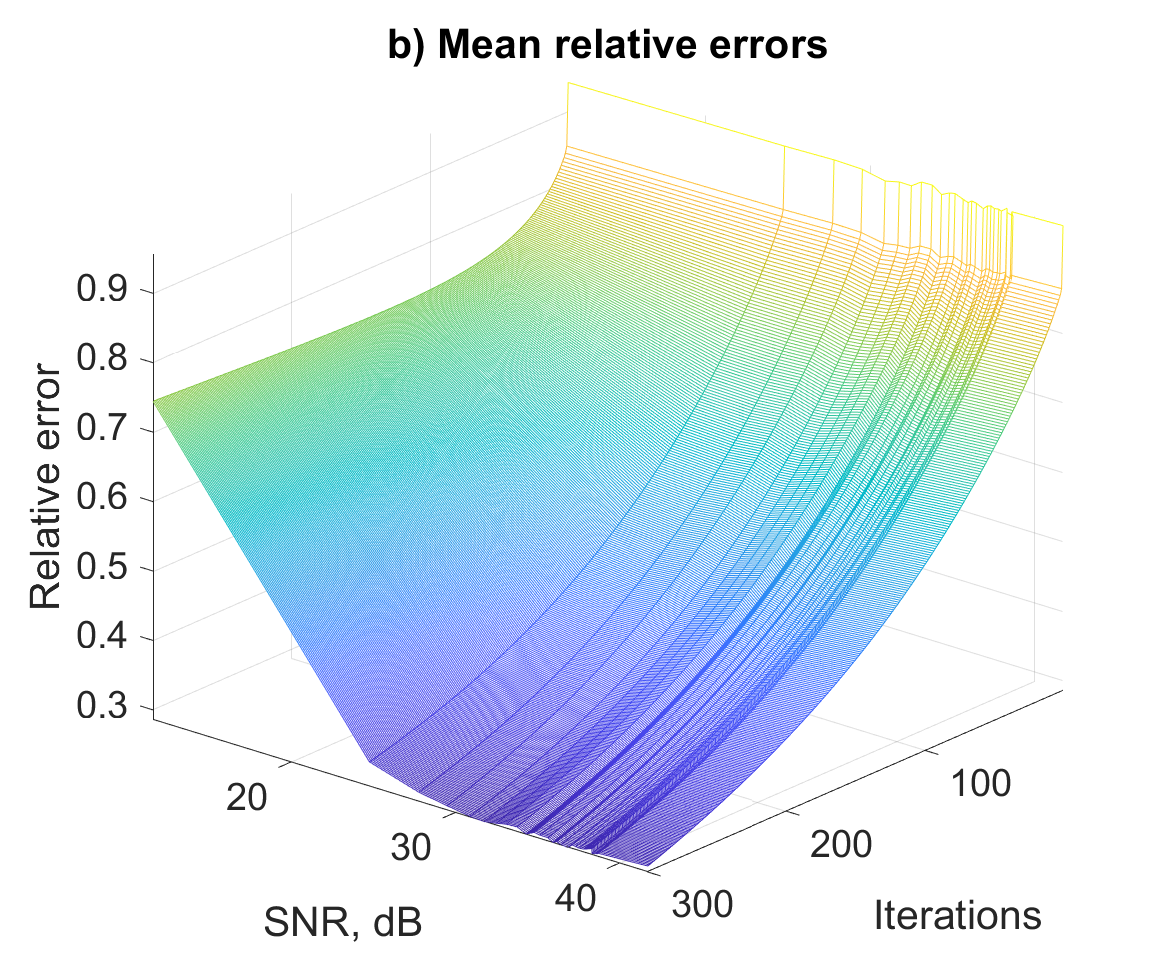} \\ b)}
\end{minipage}
 \caption{The HSPhR algorithm with disabled the Lagrange iterations and CCF filtering. The reconstruction relative error maps for the object wavefronts in the case of Poisson noise depending on: a) SNR and wavelength after 300 iterations; b) SNR and iteration number. In b) error values are represented as mean value for all wavelengths.}
\label{fig:PoissonErrors_CCFoff_Doff}
\end{figure}

 In Fig.~\ref{fig:PoissonErrors_CCFoff_Doff} we demonstrate the results obtained by the algorithm where the Lagrange iterations and CCF filtering are disabled completely. Similar as it was discussed for the Gaussian case, we may note a serious degradation in the algorithm performance with $ERROR_{rel}>~0.2$, what confirms the essential role of  both the Lagrange iterations and the CCF filtering for the algorithm performance. 
 
 
\subsection{Imaging}

\begin{figure}[h!]
\begin{minipage}[h]{.49\linewidth}
\center{\includegraphics[width=1\linewidth]{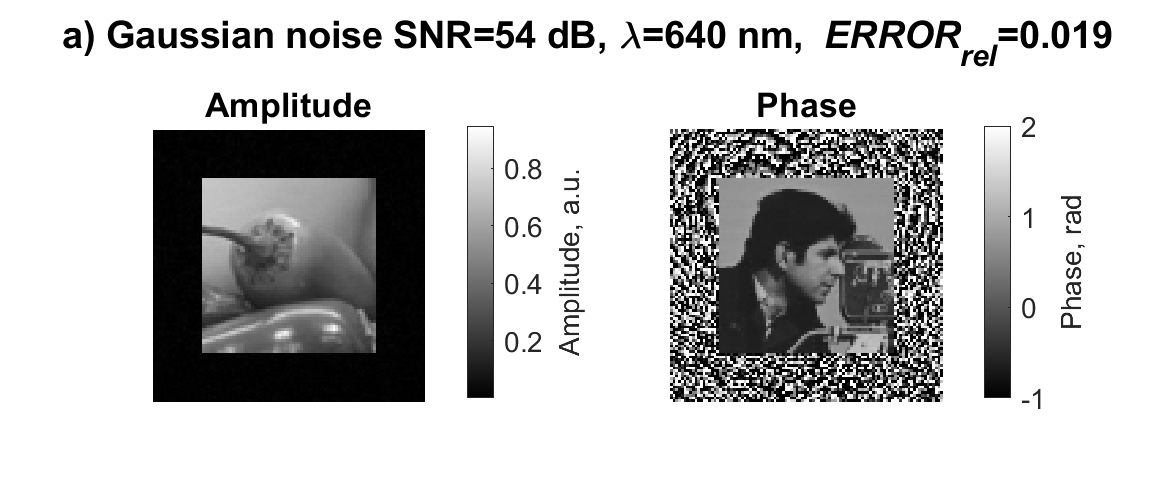}  }
\end{minipage}
\hfill
\begin{minipage}[h]{.49\linewidth}
\center{\includegraphics[width=1\linewidth]{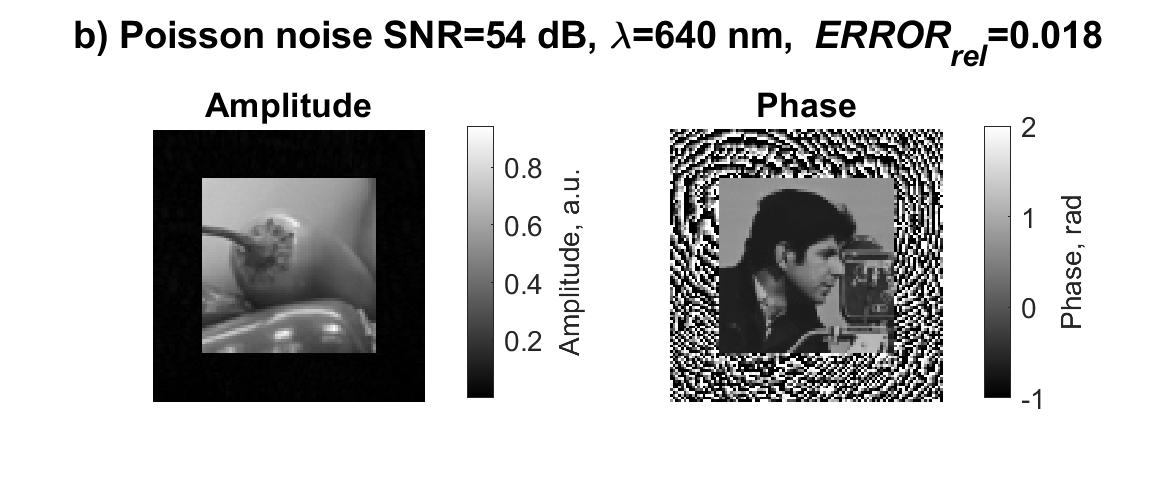}  }
\end{minipage}
 \caption{HSPhR amplitude and phase reconstructions,  nearly noiseless case (SNR=54~dB): a) Gaussian and b) Poisson versions of the algorithm;  $\lambda=640$~nm; iteration number $n=300$.}
\label{fig:Images_HSCDPR_SNR54dB}
\end{figure}

In this subsection, we  evaluate the visual quality of reconstructions. The algorithm provides the HS broadband imaging, i.e.  reconstruction of 3D complex-valued cubes. In Fig.\ref{fig:Images_HSCDPR_SNR54dB}, we show 2D images of amplitude and phase for the middle wavelength of the interval $\Lambda$,  $\lambda=640$~nm. It is a nearly noiseless case as SNR=54~dB. Fig.~\ref{fig:Images_HSCDPR_SNR54dB}(a) provides the results for the Gaussian version of the algorithm and the Gaussian noisy data, and Fig.~\ref{fig:Images_HSCDPR_SNR54dB}(b) provides the results for the Poissonian version of the algorithm and the Poissonian data. 
In all cases the images of the amplitude and phase of the high visual quality. The relative errors are low, equal to 0.019 and 0.018 for Gaussian and Poissonian data, respectively.

The images are shown in square frames  in order to emphasize that the size and location of the object support are assumed being unknown and reconstructed automatically by the HSPhR algorithm. The true image's support is used only for computation of observations produced for the zero-padded object  and is not exploited in the algorithm's iterations. For the amplitude images, these frames have nearly zero values, while the phase estimates  take random values in the frame area as the phase cannot be defined for the amplitude is equal to zero. Practically, these variations of the phase do not influence the calculation of $ERROR_{rel}$ as the amplitude estimated quite accurately in these areas and close to zero. Nevertheless, $ERROR_{rel}$ shown in the images are calculated for the central parts of the images corresponding to the true location of the image support.

\begin{figure}[h!]
\begin{minipage}[h]{.49\linewidth}
\center{\includegraphics[width=1\linewidth]{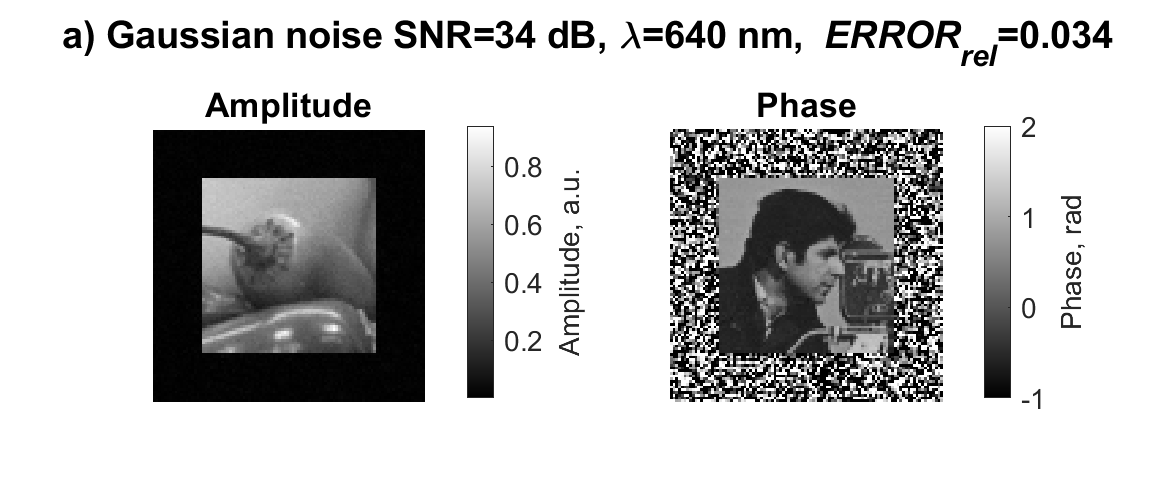}  }
\end{minipage}
\hfill
\begin{minipage}[h]{.49\linewidth}
\center{\includegraphics[width=1\linewidth]{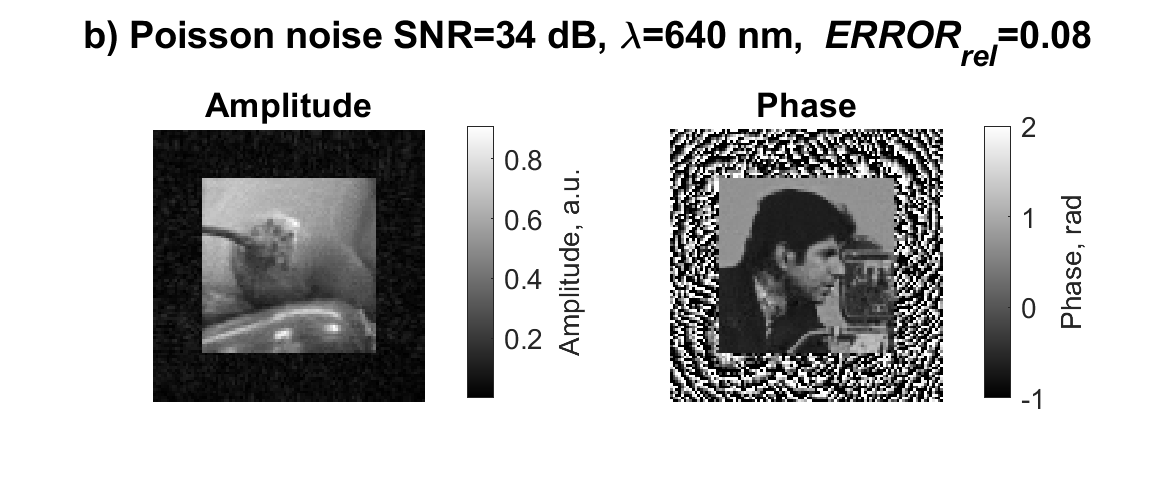}  }
\end{minipage}
 \caption{HSPhR amplitude and phase reconstructions,   noisy observations (SNR=34~dB): a) Gaussian and b) Poisson versions of the algorithm;  $\lambda=640$~nm; iteration number $n=300$.}
\label{fig:Images_HSCDPR_SNR34dB}
\end{figure}

Fig.\ref{fig:Images_HSCDPR_SNR34dB} shows the results for the noisy cases of SNR=34~dB.  With this level of SNR, the noise effects are appeared to be essential. The relative errors are much higher with 0.058 and 0.08 for Gaussian and Poissonian data, respectively. 

\begin{figure}[h!]
\begin{minipage}[h]{.49\linewidth}
\center{\includegraphics[width=1\linewidth]{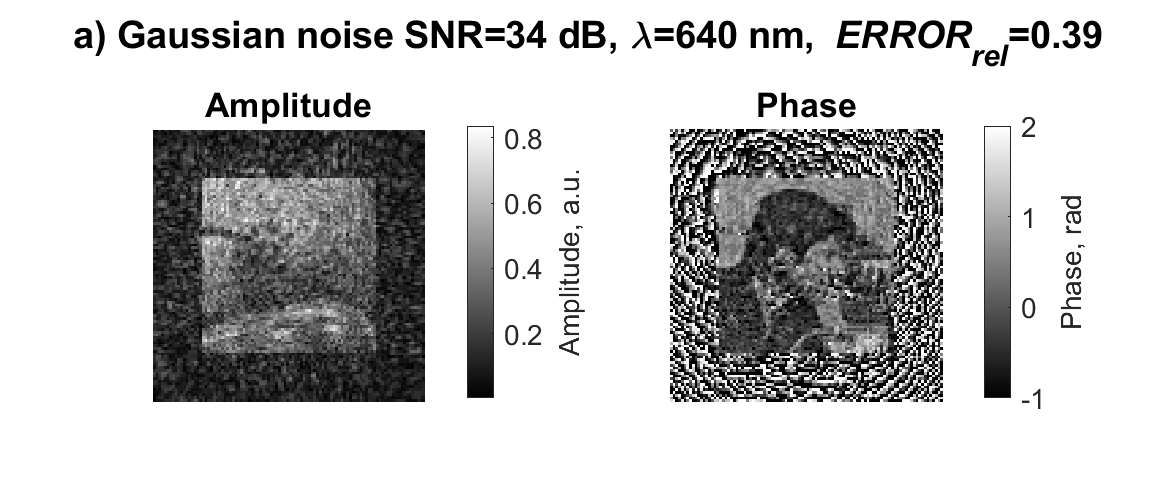}  }
\end{minipage}
\hfill
\begin{minipage}[h]{.49\linewidth}
\center{\includegraphics[width=1\linewidth]{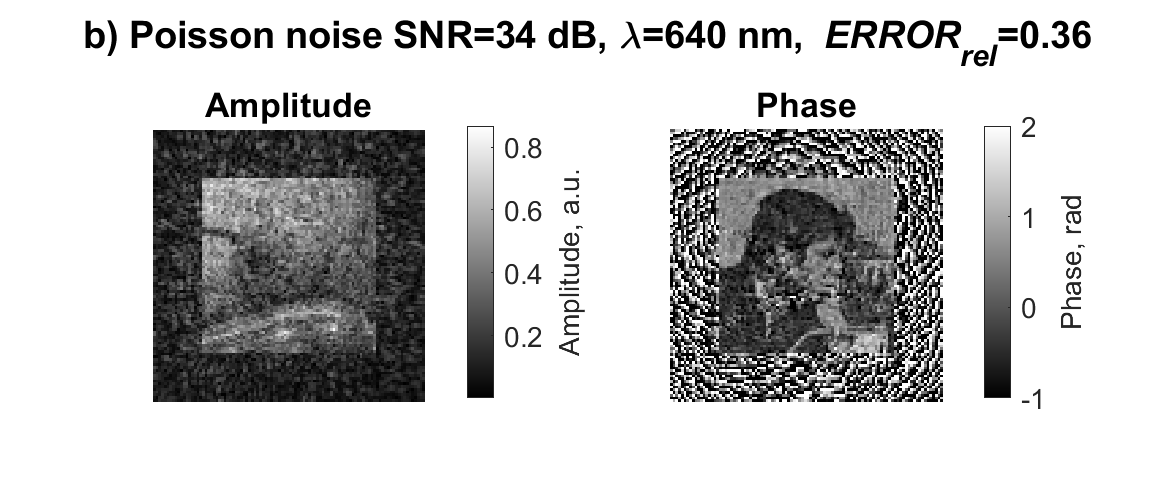}  }
\end{minipage}
 \caption{
 HSPhR amplitude and phase reconstructions, with disabled iterations on the Lagrange variables and the CCF filter; noisy observations (SNR=34~dB): a) Gaussian and b) Poisson versions of the algorithm;  $\lambda=640$~nm; iteration number $n=300$. The obtained images are very noisy, in particular, as compared with the results in Fig.\ref{fig:Images_HSCDPR_SNR34dB}.}
 
\label{fig:Images_previousAlg_SNR34_dB}
\end{figure}

In Fig.\ref{fig:Images_previousAlg_SNR34_dB} we show the  images obtained by the HSPhR algorithm for the same noisy scenario, SNR=34~dB, with the Lagrange iterations and CCF filtering disabled. The obtained amplitude/phase images are very noisy, the visual quality of imaging is low,  as compared with the results in Fig.\ref{fig:Images_HSCDPR_SNR34dB}. The relative errors take higher values equal  to 0.39 and 0.36 for Gaussian and Poissonian observations, respectively.  These experiments provide a visual and numerical confirmation of the discussed above role  of the Lagrange iterations and CCF filtering as the key components of  the HSPhR algorithm.

The complexity of the algorithm is characterized by the computational time per iteration  required for the test images, $K=6$, $T=18$. This time is equal to 0.8 sec. for calculations without CCF and equal to  21.5 sec.  for calculations with CCF. 
\textcolor{black}{ All calculations are done in MATLAB R2019b on a computer with 32 GB of RAM and CPU with a 3.40 GHz IntelR CoreTM i7-3770 processor.
We will make publicly available the MATLAB demo-code of the developed algorithm.}

\section{Conclusion} \label{sec:conclusion}
A novel formulation of the HS broadband phase retrieval problem is proposed, where both object and image formation operators are spatially and spectrally varying.  The  proposed algorithm is based on the complex domain version of
the ADMM technique. The derived Spectral Proximity
Operators  and the CCF noise suppression are important elements of this algorithm.   The SPOs  are defined for Gaussian and Poissonian  observations and calculated  solving the sets of  cubic (for Gaussian)
and quadratic (for Poissonian) algebraic equations. The ability to resolve the HS phase retrieval problem and to find the
spectral varying object components $U_{o,k}$, $k\in K$, completely depends on
 spectral properties of the spectral channels $A_{t,k}$ and  of the object $U_{o,k}$.
 The model of the object is not used in the algorithm and the modulation phase masks are calculated in advance and fixed in the algorithm iterations. The simulation tests demonstrate that the HS phase retrieval in this
formulation can be resolved with quite general modeling of
object and image formation operators. The MATLAB demo-codes of the developed algorithm are made publicly available.

\section*{Funding}

This work is a part of  the CIWIL project funded by the Technology Industries of Finland Centennial Foundation and Jane and Aatos Erkko Foundation.

%
 
\bibliographystyle{IEEEtran}

\end{document}